# PROGRESS WITH PXIE MEBT CHOPPER

V. Lebedev[#], A. Chen, R. Pasquinelli, D. Peterson, G. Saewert, A. Shemyakin, D. Sun, M. Wendt

Fermilab*, Batavia, IL 60510, USA

T. Tang, SLAC, Menlo Park, CA 94025

*Abstract*

A capability to provide a large variety of bunch patterns is crucial for the concept of the Project X [1] serving MW-range beam to several experiments simultaneously. This capability will be realized by the Medium Energy Beam Transport's (MEBT) chopping system that will divert 80% of all bunches of the initially 5mA, 2.1 MeV CW 162.5 MHz beam to an absorber according to a pre-programmed bunch-by-bunch selection. Being considered one of the most challenging components, the chopping system will be tested at the Project X Injector Experiment (PXIE) [2] facility that will be built at Fermilab as a prototype of the Project X front end. The bunch deflection will be made by two identical sets of travelling-wave kickers working in sync. Presently, two versions of the kickers are being investigated: a helical 200 Ohm structure with a switching-type 500 V driver and a planar 50 Ohm structure with a linear ±250 V amplifier. This paper will describe the chopping system scheme and functional specifications for the kickers, present results of electromagnetic measurements of the models, discuss possible driver schemes, and show a conceptual mechanical design.

## SCHEME AND REQUIREMENTS

The PXIE MEBT chopping scheme is shown in Fig.1. Two 50cm–long kickers are separated by 180º phase advance and have opposite direction of the electric field during passage of the same bunch (similar to the logic of Ref.[3]) so that the kicks to the bunch are summed.

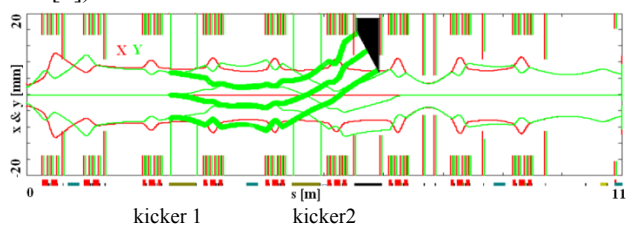

Figure 1: Scheme of MEBT optics [4] and the beam envelope. The thin lines are the central trajectory and 3σ envelope of the passing beam, and the thick lines are the Y envelope of the chopped-out beam.

The kicker field is generated by applying voltage of opposite polarity to two sets of flat electrodes above and below the beam. With the typical beam transverse size of $\sigma_{x,y} \approx 2$ mm, the vertical gap between electrodes is chosen to be 16 mm. The aperture in the vicinity of the kickers is limited by pairs (one at each end) of protection electrodes with the vertical gap of 13 mm. We plan to interrupt the beam if the lost current to any of these electrodes exceeds ~10 μA to protect the kickers.

A voltage is applied to the kicker electrodes at passage of each bunch. To provide the separation shown in Fig.1, the electrode voltage, for example, at the top electrodes of the first kicker is -250 V for the passing bunches and +250 V for the bunches to be removed. During the bunch duration, $6\sigma_z \approx 1.3$ ns, the voltage should stay flat within ±25V for passing bunches and should be ≥250 V for the chopped bunches. This voltage is calculated for an ideal kicker, and for the present kicker designs the driver should provide ~10% higher value. Removal of at least 80% of bunches means that the maximum switching rate between pass/remove states, averaged over the typical ~1 μs period, is 162.5/5=32.5MHz. The kicker phase velocity should be matched to the beam speed of 20 mm/ns. The flange-to-flange length of the kicker assembly is 650 mm. The vacuum in the kicker vicinity with a full-power beam is expected to be $2 \cdot 10^{-7}$ Torr.

Two versions of the kicker and driver, referred below by their impedance, 50 and 200 Ohm, are being developed.

## 50 OHM VERSION

The 50-Ohm kicker consists of two identical structures, which mechanical design is shown in Fig.2. The beam is deflected by voltage applied to planar electrodes connected in vacuum by coaxial cables with the length providing necessary delays, similar to the design used at LAMPF [5]. The 50cm-long structure has 25 copper plates of 14.6 mm length (along the beam trajectory), 50mm width and 1 mm thickness. The total power loss in the electrodes induced by the electromagnetic signal is expected to be ~110 W per structure, and the structure is required to withstand additional 40 W of the beam loss. Cooling of the Teflon-insulated cables is provided by clamps, which, in turn, are cooled by water flowing through the channels in the structure.

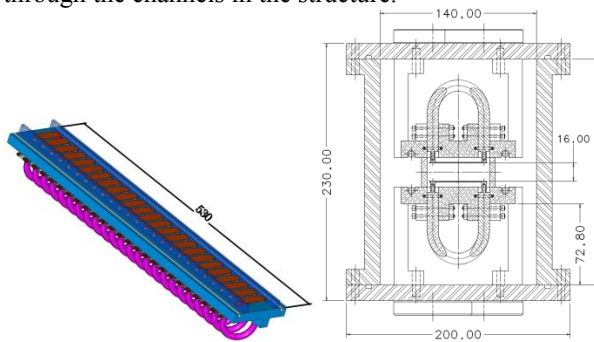

Figure 2: Mechanical schematic of the 50-Ohm kicker structure and the cross section of the kicker assembly.

Results of simulations and measurements of the electromagnetic model containing 6 electrodes are shown in Fig. 3. The expected attenuation in the full-length (50cm) structure is ~0.5 dB, and the predicted signal distortion is low enough to provide the specified parameters.

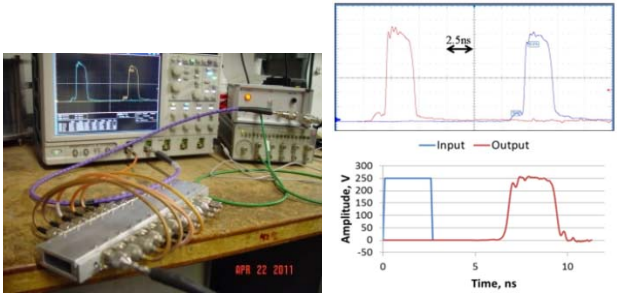

Figure 3: The six-electrode electromagnetic model of the 50-Ohm kicker structure and results of its measurements (top right). The shape of the output signal (violet) is very close to the input (reddish). The bottom right plot shows the simulated distortion of the output signal (red) in a 8-electrode structure after injecting a rectangular pulse (blue).

The kicker will be driven by a linear amplifier. To decrease the lower frequency content of the output signal, the 6.15ns pulse affecting a single bunch is formed as shown in Fig.4c.

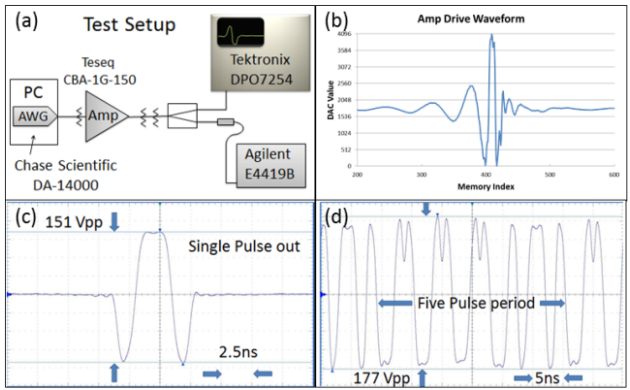

Figure 4: Test of the CBA 1G-150 amplifier with pre-distortion. (a) scheme of the test; (b) pre-distorted input signal and (c) corresponding output signal for a single pulse; (d) output for a CW pattern, corresponding to removal of four consecutive bunches followed by a one-bunch passage.

After testing amplifiers from several companies, performance of a commercially available CBA 1G-150, a 150W, 0.05-1 GHz amplifier from Teseq AG, Switzerland [6] was found satisfactory. To compensate for a significant low frequency phase distortion, the following procedure was proposed and tested:

1. The amplifier impulse response was measured using a single 250 ps pulse from the arbitrary waveform generator (AWG), and the complex gain function of the amplifier was calculated.

2. The desired waveform is presented as a sum of individual pulses shaped as in Fig.4c with corresponding polarity. Note that the bunch pattern is expected to be repeated at 1μs intervals.

3. The waveform was chosen so that its spectrum would be inside of the amplifier bandwidth.

4. The driving signal for the amplifier (Fig.4a) was obtained as inverse Fourier transform for ratio of the waveform spectrum to the amplifier gain. Then, the obtained waveform was fed into the arbitrary waveform generator [7].

The test results are shown in Fig.4b,c,d. The pulse shape was very good for single pulses and deteriorated when the average power was approaching the power limit of the amplifier. In the future we plan to correct this distortion using iterative algorithm for the input pulse generation. The maximum signal amplitude with the shape satisfying the specification was 240 V peak to peak.

According to the company, it has a commercially available 1 kW amplifier with similar characteristics that should provide the voltage amplitude required for the chopper.

## 200 OHM VERSION

The 200 Ohm kicker consists of two helical structures (Figure 5). A helical deflector was used at LAMPF [8] (with a significantly slower rise time) and in Tektronix oscilloscope 7104. This helix is a flat wire wound around a 28.6 mm OD copper tube that serves as the signal return. The wire is suspended 4.8 mm above the tube by four ceramic spacers. The combination of this space and the 8.5mm helix pitch results in 200 Ohm characteristic impedance. The deflecting field is formed by flat copper 5.6x20x0.5 mm electrodes soldered to each turn of the helix.

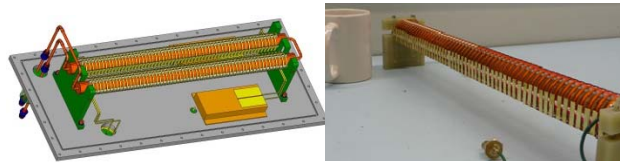

Figure 5: Mechanical drawing of 200 Ohm dual-helix kicker and photo of a single 200 Ohm helix prototype.

While heating from pulsing is estimated to be low (< 10 W), the structure is rated for removal of 40W deposited by beam tails. The heat is removed mainly through the ceramic spacers contacting the water-cooled central tube.

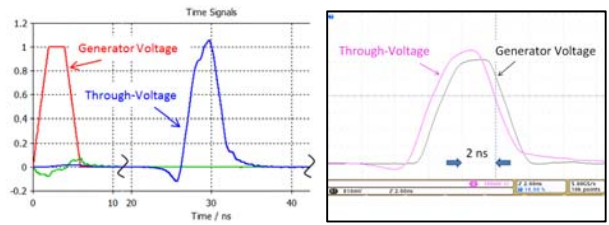

Figure 6: Simulation results of the helix kicker (left) and the prototype response (right). The traces at the right plot are time shifted.

Simulations of the kicker structure compare well with prototype measurements (Fig. 6). Distortion during the propagation through the 200 Ohm helix is within specifications.

The 200 Ohm kickers are intended to be driven by broadband, DC coupled switches in push-pull configuration. The switching waveform pattern will be made to vary depending on the timing requirements of the experiments receiving beam. This design reduces the operating switching rates resulting in lower driver circuit switching losses as well as lower power dissipated in the helical structure.

The switch-driver is being developed in two versions. The first version is based on GaN multi-FET cascode scheme where five 200 V rated FETs [9] turn on together and share the voltage. Fig. 7a shows the circuit topology used for this switch, and Fig. 7b presents scope traces of preliminary tests of the individual transistors sharing voltage while switching. The rise time of the output signal, 3 ns (10-90%), is close to the requirement.

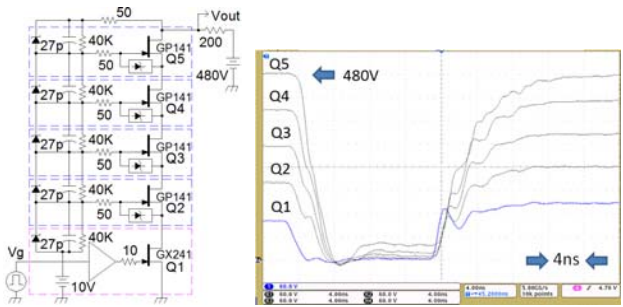

Figure 7: Simplified schematic of a multi-FET cascode switch and scope traces of the five FET drain voltages.

The second version is based on a 500V ultra-fast hybrid MOSFET/Driver switching module (HSM) [9], which is being developed at SLAC. The HSM consists of a pair of complementary MOSFETs arranged in totem pole configuration. The MOSFETs are acquired in bare die form and are attached to the circuit board using low-inductance assembly method to achieve ultra fast switching. A switching time of less than 1ns is observed in the simulation, which will be experimentally verified once the HSMs are fully fabricated and assembled.

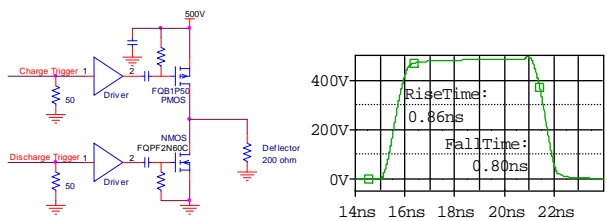

Figure 8: Schematic of HSM switch and simulation results.

## DISCUSSION AMD PLANS

Presently the schemes of the kicker and driver described above are being pursued in parallel. On one hand, the 50-Ohm version employs a proven kicker technology, commercially available feedthroughs, cables, and loads, as well as is capable of using off-the-shelf amplifiers (with a special procedure of pre-distortion). On the other hand, the 200 Ohm scheme has a potential to be of a lower cost, and its DC coupling may be beneficial for forming the bunch patterns with long periods. So far, development of both versions does not show any clear showstoppers, and the final technology choice will be made later.

Both kicker structures are at the stage of mechanical design of vacuum-compatible prototypes. Manufacturing and following tests (electromagnetic, vacuum, and thermal) are scheduled for CY 2012.

## ACKNOWLEDGMENT

The authors acknowledge the insightful suggestion by B. Chase (Fermilab) for the helical traveling wave structure concept; the engineering assistance of D. Frolov, an intern from Kuban State University, Russia, for his work on the cascode driver, and the proposal for protection electrodes from A. Aleksandrov.